 \documentclass[final,5p,times,twocolumn,authoryear]{elsarticle}

\usepackage{txfonts}
\usepackage{pdflscape}
\usepackage{enumitem}
\usepackage{framed}
\usepackage{pgfgantt,rotating}
\usepackage{multirow}
\usepackage{blindtext}
\usepackage[T1]{fontenc}
\usepackage{ae,aecompl}
\usepackage{graphicx}	
\usepackage{amsmath}	
\usepackage{rotating}
\usepackage{listings}
\usepackage{threeparttable}
\usepackage{wasysym,url}

\usepackage{amssymb}
\usepackage{lipsum}

\journal{High Energy Astrophysics}

\begin{document}

\begin{frontmatter}



\title{Hunting for extreme high-energy-peaked BL Lacs: \\ 
Rare to find and difficult to classify}


\author[1]{Loredana Bassani}
\ead{loredana.bassani@inaf.it}
\affiliation[1]{organization={INAF -- Osservatorio di Astrofisica e Scienza dello Spazio},
            addressline={Via Piero Gobetti 101}, 
            city={Bologna},
            postcode={40129}, 
            state={Italy}}


     \author[1]{Raffaella Landi}
    \ead{raffaella.landi@inaf.it}

 \author[1,2]{and Nicola Masetti}
    \ead{nicola.masetti@inaf.it}

\affiliation[2]{organization={Instituto de Astrofisica, Universidad Andres Bello},
            addressline={Av. Fernandez Concha 700}, 
            city={Las Condes, Santiago},
            state={Chile}}

%
%

\begin{abstract}
We explore the possible existence of a new population of BL Lacs, called ultra extreme high-energy-peaked BL Lacs (UEHBLs),
whose synchrotron emission component peaks in the MeV band. In particular, we analysed \emph{Swift}/XRT follow-up observations of a set of 10 hard X-ray sources from the \emph{Swift}/BAT catalogues which were suggested to represent the first observational
hints of this new blazar population on the basis of their spectral properties.

We find that 6 of these candidate UEHBLs can be classified as line-emitting active galactic nuclei (AGNs), 4 of which are of type 2 and X-ray absorbed objects, while 2 are type 1 AGN and X-ray unabsorbed sources. 
In one more case, we find that the hard X-ray emission is probably the contribution of two line-emitting AGN, i.e. a Seyfert of type 1.9 and another of type 2. All these 7 classifications exclude the possibility that these hard X-ray sources are extreme BL Lacs. 

Of the remaining 3 UEHBL candidates, only 2 objects are found to have a spectral energy distribution (SED) compatible with those expected by 
MeV-peaked BL Lacs: however, in one of these 2 cases, an alternative Galactic association is also possible. The third source is instead likely associated with a more classical quasi-stellar object.

Overall, we find that UEHBLs are extremely rare to find and probably very difficult to classify. Nevertheless, they represent a viable alternative classification for hard X-ray sources displaying no obvious or very weak counterparts.

\end{abstract}




\end{frontmatter}




\section{Introduction}
\label{introduction}

Among the blazar population extreme objects are probably the most intriguing, as they allow one to study particle acceleration and radiative processes in the most extreme environments and to explore links with cosmic rays and neutrino astrophysics. So far, two types of extreme blazars have been observed: those showing a synchrotron peak above 1 keV and up to 100 keV and those showing an Inverse Compton peak above 1 TeV and up to 10 TeV. Overall, they are well studied objects as their multi-wavelength properties and spectral energy distributions (SED) have been extensively analyzed and discussed in the literature (see for examples, \citealt{Costamante2001}, \citealt{Biteau2020}, \citealt{Acciari2020}, \citealt{MAGICCollaboration2024}).

Recently, however, \citet{Sciaccaluga2025} 
suggested the potential existence of a new population of even more extreme objects, called ultra extreme high-energy-peaked BL Lacs (UEHBLs), whose synchrotron emission component peaks in the MeV band, while the Inverse Compton component is expected in the GeV/TeV region according to their modeling of these sources SED. However, due to severe Klein-Nishina suppression of Inverse Compton scattering, these UEHBLs would be undetectable with current GeV (\emph{Fermi}) and future TeV 
(i.e., Cherenkov Telescope Array Observatory, CTAO, \citealt{CTA2019}) facilities, although they represent ideal targets for MeV missions and could have already been discovered by hard X-ray telescopes like \emph{Swift}/BAT \citep{Barthelmy2005}, and \emph{INTEGRAL}/IBIS \citep{Ubertini2003}.

Based on current data, these authors also proposed a sample of 10 hard X-ray sources extracted from the \emph{Swift}/BAT 157-month survey catalogue \citep{Lien2025}, potentially representing the first observational hints of this population. 

These UEHBLs were chosen following a set of criteria: having a relatively bright and flat spectrum in the hard X-ray energy band (flux above $10^{-11}$ erg cm$^{-2}$ s$^{-1}$ and photon index $\Gamma < 1.5$) to ensure detectability in the 
MeV regime, a clear extragalactic origin (defined by a Galactic latitude |b| > 10$^{\circ}$), and finally showing no obvious counterpart in the soft X-ray and optical wavebands, as UEHBLs are expected to be extremely weak at these frequencies according to the SED modeling. 

Confirmation of their nature as extreme BL Lacs is very important as it would not only push the blazar sequence beyond its current limits, but it would also allow the characterisation of the most extreme accelerators among blazars, specifically, and AGN population more generally.

Here, we present follow-up X-ray observations, acquired with the
X-ray Telescope (XRT, 0.3--10 keV, \citealt{Burrows2005}) on board the Neil Gehrels \emph{Swift} Observatory \citep{Gehrels2004}, of all of these 10 candidates which allow us to find likely counterparts for most sources, study their X-ray properties and, ultimately, confirm or reject their classification as UEHBLs. As a results of this work, we are able to exclude from the list 7 objects and focalise instead on a small set of 3 objects, which possibly satisfy the SED criteria to be ultra extreme BL Lacs. As a byproduct of this work we also identify and classify 8 \emph{Swift}/BAT sources, including one associated with a \emph{Fermi}/LAT object; for the remaining 2 sources the association remains speculative.

\section{Data reduction and analysis}
X-ray pointings for all but 2 sources in the sample were already available in data archives; for the two remaining sources, SWIFT J0045.9$+$3931 and SWIFT J0243.2--0553, we triggered Target of Opportunity observations (ToOs) with \emph{Swift}/XRT, thus allowing us to cover the entire sample with reasonable exposures.

XRT data reduction was performed using the standard data pipeline package {\sc xrtpipeline} v.~0.13.7, to produce screened event files. All data were extracted only in the Photon Counting (PC) mode \citep{Hill2004}, adopting the standard grade filtering (0--12 for PC) according to the XRT nomenclature.

In case of multiple observations, we summed together all the available XRT pointings using {\sc XSELECT} v.~2.5c to enhance the signal-to-noise ratio and thus facilitate the detection of candidate counterparts. As a following step, we analysed the XRT images in the 0.3--10 keV energy band by means of {\sc XIMAGE} v.~4.5.1 in search of X-ray detections within or nearby the BAT error circles.

The position and the associated positional error of the X-ray detections were estimated by using the task {\sc xrtmkarf} v.~0.6.4.

In the XRT images (Figures from 1 to 4), the continuum and dotted black circles depict the 90\% and 99\% BAT positional uncertainties, respectively, as reported in the 157-month \emph{Swift}/BAT catalogue \citep{Lien2025}.

To visualise better the X-ray counterparts, we smoothed the images, therefore the presence of grains and/or features inside the XRT field of view is undoubtedly spurious.

Most counterparts were chosen based on their X-ray flux brightness and spectral hardness (i.e. a detection above few keV), as these two parameters enhance the probability of a true association with a hard X-ray emitter \citep{Stephen2006}.

In Table~\ref{tab1} we list the set of UEHBL candidates suggested by \citet{Sciaccaluga2025} together with their coordinates and relative positional uncertainty.
For each of these X-ray emitters, we then report the coordinates and relative uncertainties (at 90\% confidence level) of the X-ray sources detected by XRT within or nearby the BAT error circles, their count rate in both the 0.3--10 and 3--10 keV energy range, the number of XRT observations used for the current analysis and the total on-source exposure time.
Figures from 1 to 4 provide the 0.3--10 keV XRT images of the sky region around 4 \emph{Swift}/BAT sources that need some clarifications regarding their likely counterpart(s).

\begin{table*}
\begin{center}
\footnotesize
\caption{\emph{Swift}/BAT position of the set of UEHBLs selected by \citet{Sciaccaluga2025} and locations of the objects detected by XRT, within the BAT positional uncertainties, with relative count rates in the 0.3--10 and 3--10 keV energy range, their distance to the best-fit BAT source position, the number of X-ray observations analysed and the total on-source exposure time. The XRT error radii are given at 90\% confidence level. See text for details.}
\label{tab1}
\scriptsize
\begin{tabular}{lccccccc}
\hline
\hline
  &      &   &   &   & &  & \\
XRT source  &     R.A.     &     Dec.   &   error   &  \multicolumn{2}{c}{Count rate} & Distance  &  N. obs/Total expo$^{a}$    \\
            &              &            &          &   (0.3--10 keV)  &  (3--10 keV) &  &  \\
  &   (J2000) &  (J2000) &   (arcsec)  &  (10$^{-3}$ counts s$^{-1}$) & (10$^{-3}$ counts s$^{-1}$) & (arcmin)   &   (s)   \\
\hline
\hline
   &  &  &   & &  & & \\
\multicolumn{8}{c}{\textbf{SWIFT J0007.8--4133} (R.A.(J2000) = $00^{\rm h}07^{\rm m}49^{\rm s}.08$,
Dec.(J2000) = $-$$41^\circ21^{\prime}21^{\prime \prime}.20$, err(90\%) = 5$^{\prime}$.75)}  \\
   & &     &  & &   &   & \\ 
\#1 & $00^{\rm h}07^{\rm m}06^{\rm s}.68$ & $-$$41^\circ21^{\prime}19^{\prime \prime}.84$ & 3.62 & $39.90\pm1.60$ &  $7.75\pm0.70$  & 7.96  &  12/22935\\
   & &     &   &   &  & & \\
   \hline
      & &     &      & & &  & \\
      \multicolumn{8}{c}{\textbf{SWIFT J0045.9$+$3931} (R.A.(J2000) = $00^{\rm h}46^{\rm m}02^{\rm s}.21$,
Dec.(J2000) = $+$$39^\circ31^{\prime}51^{\prime \prime}.60$, err(90\%) = 5$^{\prime}$.91)}  \\
  & &     & &      &  &  & \\
\#1  &  $00^{\rm h}45^{\rm m}58^{\rm s}.17$ & $+$$39^\circ33^{\prime}50^{\prime \prime}.06$ & 5.50 & $9.72\pm1.80$  &  $4.79\pm1.30$  &
2.12   &  2/3606 \\
& &      &   & & & & \\
   \hline
       &     &   & & &   &   & \\
      \multicolumn{8}{c}{\textbf{SWIFT J0106.1$+$4818} (R.A.(J2000) = $01^{\rm h}06^{\rm m}05^{\rm s}.57$,
Dec.(J2000) = $+$$48^\circ17^{\prime}41^{\prime \prime}.60$, err(90\%) = 5$^{\prime}$.63)}  \\
  & &     &   &   &   &  &  \\
\#1  & $01^{\rm h}05^{\rm m}50^{\rm s}.11$ & $+$$48^\circ19^{\prime}04^{\prime \prime}.21$ & 4.19 & $17.40\pm1.80$ &  $10.50\pm1.40$ & 2.92  &
1/6895\\
&     &   &   &  &  &   &\\
   \hline
      &     &   &   &  & &  & \\
      \multicolumn{8}{c}{\textbf{SWIFT J0243.2--0553} (R.A.(J2000) = = $02^{\rm h}43^{\rm m}13^{\rm s}.03$,1
Dec.(J2000) = $-$$05^\circ53^{\prime}12^{\prime \prime}.10$, err(90\%) = 5$^{\prime}$.48)} \\
  &  &     &   &    & &  &\\
\#1 & $02^{\rm h}43^{\rm m}12^{\rm s}.70$ & $-$$05^\circ50^{\prime}58^{\prime \prime}.13$ & 5.76 & $32.10\pm6.80$ & --  & 2.23   &
1/814 \\
&  &     &   &   &  & & \\
   \hline
      & &     &    &  & &  & \\
      \multicolumn{8}{c}{\textbf{SWIFT J0449.3$+$6356} (R.A.(J2000) = $04^{\rm h}49^{\rm m}18^{\rm s}.77$,
Dec.(J2000) = $+$$63^\circ56^{\prime}00^{\prime \prime}.20$, err(90\%) = 4$^{\prime}$.72)} \\
  & &     &   &    &  &   & \\
\#1 & $04^{\rm h}48^{\rm m}09^{\rm s}.60$ & $+$$63^\circ54^{\prime}50^{\prime \prime}.30$ & 6.4 & $3.42\pm1.00$  &   -- & 7.69  &
8/6491  \\
 \#2 &  $04^{\rm h}48^{\rm m}18^{\rm s}.90$ & $+$$64^\circ01^{\prime}13^{\prime \prime}.70$ & 4.9 & $8.22\pm1.80$  &   --  & 8.39 &    \\
& &     &   &   &  & &  \\
   \hline
      & &     &   &   & &   & \\
      \multicolumn{8}{c}{\textbf{SWIFT J0656.0--6560} (R.A.(J2000) = $06^{\rm h}55^{\rm m}57^{\rm s}.36$,
Dec.(J2000) = $-$$65^\circ35^{\prime}53^{\prime \prime}.20$, err(90\%) = 5$^{\prime}.50$)} \\
  & &     &   & &    &   & \\
\#1 & $06^{\rm h}56^{\rm m}29^{\rm s}.95$ & $-$$65^\circ33^{\prime}39^{\prime \prime}.12$ & 3.57 & $86.20\pm2.60$ &  $11.90\pm0.85$ & 4.04 &  7/19002 \\
&      &   & &  &  & & \\
   \hline
       &   &  &   &  & &   & \\
      \multicolumn{8}{c}{\textbf{SWIFT J0722.5$+$2121} (R.A.(J2000) = $07^{\rm h}22^{\rm m}40^{\rm s}.32$,
Dec.(J2000) = $+$$21^\circ26^{\prime}26^{\prime \prime}.90$, err(90\%) = 6$^{\prime}.61$)}  \\
  & &     & &  &   &  &  \\
\#1  & $07^{\rm h}22^{\rm m}23^{\rm s}.73$ & $+$$21^\circ25^{\prime}05^{\prime \prime}.70$ & 5.20 & $1.92\pm0.48$  &  -- & 4.09 &
5/12637  \\
 &     &   &   &  & &  & \\
   \hline
      &     &   &   &  & &  & \\
      \multicolumn{8}{c}{\textbf{SWIFT J1026.3$+$4536} (R.A.(J2000) = $10^{\rm h}26^{\rm m}20^{\rm s}.64$,
Dec.(J2000) = $+$$45^\circ35^{\prime}57^{\prime \prime}.10$), err(90\%) = 4$^{\prime}.95$)}  \\
  & &     & &  &   &  &  \\
\#1 & $10^{\rm h}26^{\rm m}19^{\rm s}.11$ & $+$$45^\circ34^{\prime}43^{\prime \prime}.29$ & 3.89 & $7.98\pm0.71$ & $2.79\pm0.39$  &  1.26  &  30/22780 \\
& &     &   &  &  & &  \\
   \hline
      & &  &   &   &   &   & \\
      \multicolumn{8}{c}{\textbf{SWIFT J1334.1--3842} (R.A.(J2000) = $13^{\rm h}33^{\rm m}44^{\rm s}.88$,
Dec.(J2000) = $-$$38^\circ27^{\prime}00^{\prime \prime}.40$, err(90\%) = 5$^{\prime}.22$)}\\
  & &     &  & &   &  &  \\
\#1 & $13^{\rm h}33^{\rm m}58^{\rm s}.59$ & $-$$38^\circ24^{\prime}54^{\prime \prime}.20$ & 4.67 & $21.50\pm3.30$ & $11.60\pm2.20$ &  3.41  &  2/2672 \\
\#2 & $13^{\rm h}33^{\rm m}53^{\rm s}.76$ & $-$$38^\circ25^{\prime}48^{\prime \prime}.66$ & 5.00 & $7.56\pm2.00$ &  $4.03\pm1.40$ &  2.11  &   \\
   &  &  &  &  &  &  &  \\
   \hline
      & &  &   &   &  &    & \\
      \multicolumn{8}{c}{\textbf{SWIFT J1949.7--3636} (R.A.(J2000) = $19^{\rm h}49^{\rm m}44^{\rm s}.64$,
Dec.(J2000) = $-$$36^\circ35^{\prime}52^{\prime \prime}.40$, err(90\%) = 5$^{\prime}.21$)}\\
   &   &  &   &   &  &  &\\
\#1 & $19^{\rm h}49^{\rm m}50^{\rm s}.99$ & $-$$36^\circ35^{\prime}23^{\prime \prime}.54$ & 3.72 & $30.10\pm1.80$ &  $22.10\pm1.40$ & 1.36  &   14/13790  \\
   &  & & &  &  &  & \\
\hline
\hline
\end{tabular}
\begin{list}{}{}
\item $^{a}$: On-source exposure.
\end{list}
\end{center}
\end{table*}

For each object, source events were extracted from the corresponding event file within a circular region with a radius of 20 pixels (1 pixel corresponding to 2.36 arcseconds) centred on the source position, while background events were extracted from a source-free region close to the X-ray source of interest. Then, the source spectra were extracted using the {\sc XSELECT} v.~2.5c software and generally binned using {\sc grppha} to 20 counts per energy bin so that the $\chi^{2}$ statistic could be applied. For objects with fewer counts (typically lower than 50), data were binned to 1 count per energy bin and the Cash statistic (Cash 1979) was adopted. We used version v.~016 of the response matrices and created individual ancillary response files \emph{arf} using {\sc xrtmkarf} v.~0.6.3. The spectral analysis was performed using {\sc XSPEC} v.~12.15.0d \citep{Arnaud1996}.

To broadly characterise each source spectrum we adopted a basic model consisting of a simple power law passing through Galactic absorption in the source direction \citep{Kalberla2005}. 
In those cases where the power law returns a flat spectral index, we introduce extra absorption intrinsic to the source. If the statistics does not allow us to estimate the uncertainty of both photon index and column density, the first is set to 1.8, taken as typical of an AGN spectrum.

The results of this spectral analysis are reported in Table~\ref{tab2}.

\section{From X-ray counterparts to source classification and properties}

In most cases, the BAT/XRT association is pretty straightforward since a single bright and hard X-ray counterpart is immediately evident in the XRT images. In the remaining cases, the association is not so straightforward as expected and some more in-depth analysis was required.

All likely or potential X-ray counterparts are summarised in Table~\ref{tab2}, where a set of information is reported including optical identification and classification, source redshift and relative reference, as well as Galactic and intrinsic X-ray column density, photon index, and 2--10 keV X-ray flux.

In the following, we briefly describe simple cases where a single and hard counterpart was found.

 \subsection{Simple cases}

One such case is that of SWIFT J0045.9$+$3931, as the only likely association is LEDA 2151989, a Seyfert galaxy of the local Universe (see Appendix).
The source is characterised by radio emission being detected by the LOw-Frequency ARray (LOFAR, \citealt{vanhaarlem2013}) at 0.144 GHz with a flux of 11.3 mJy \citep{Shimwell2026}
and by the Australian Square Kilometre Array Pathfinder (ASKAP) at 1.3/1.6 GHz with a flux of 3.4/2.8 mJy (\citealt{Duchesne2024}; \citealt{Duchesne2025}), while its far-infrared colours ($W1-W2 = 0.68$ and $W2-W3=2.69$) from the Wide-field Infrared Survey Explorer (\emph{WISE}, \citealt{Wright2010}) are
typical of active galaxies. The optical spectrum allows the classification of the source as a type 2 AGN and the X-ray spectrum, although of low statistical quality, reflects this classification as the source is mildly absorbed.
All together, we conclude that this source, due to its classification and properties, should be erased from the list of extreme BL Lac candidates.\\

The case of SWIFT J0106.1$+$4818 is also straightforward as the source is associated with the AGN candidate ICRF J010549.9$+$481903, probably a jetted object which has no measured redshift so far. As this source is one of the small number of UEHBL candidates remaining after our analysis, it will be discussed more in details in Section 4.\\

SWIFT J0243.2--0553 is interesting as it is the only source in the sample with a counterpart in the 4FGL-DR4 \emph{Fermi} catalogue \citep{Abdollahi2022},
i.e. with 4FGL J0243.2--0550, itself associated with the flat spectrum radio quasar (FSRQ) PKS 0240--060 at redshift $z=1.8$. PKS 0240--060 is a high-redshift object with broad optical emission lines in its spectrum (see \citealt{Hook2003}) and strong and flat radio emission, hence it is classified as a FSRQ. This may well be one of those cases discussed by \citet{Sciaccaluga2025}, where the \emph{Swift}/BAT source has a high-energy MeV peak associated with the Compton emission component and not with the synchrotron one. However, no soft X-ray data were available for this sky region and the \emph{Fermi}/FSRQ association was purely done on the basis of the source radio data and optical classification. 
To confirm the association between \emph{Fermi}/BAT and this FSRQ, we triggered an XRT ToO observation of this sky region and indeed found that the quasar is the only counterpart to both hard X-ray and gamma-ray sources. Due to its properties, also this source can be excluded from the list of candidates UEHBLs.\\

SWIFT J0656.0--6560 (also IGR J06569--6534) is also a simple case as the source is associated with FRL 265, an active galaxy of the local Universe displaying a classical type 1 Seyfert spectrum. This hard X-ray source has been extensively discussed in the literature before (\citealt{Pal2026}; \citealt{Caglar2023}; \citealt{Malizia2020})
and will not be analysed further here; however, given its classification, this source can be taken out from the list of UEHBL candidates proposed by \citet{Sciaccaluga2025}.\\

Another simple case is that of SWIFT J1026.3$+$4536, whose counterpart is Z 240--42, a radio galaxy of unknown classification located at redshift $z= 0.02689$. Z 240--42 is part of a double system, the companion being LEDA 2267439 at redshift $z=0.02656$. It was suggested to be an AGN by \citet{Cavuoti2024} and \citet{Zaw2019}, and classified as such by \citet{Condon2019} on the basis of four indicators which use radio and infrared data.

The source is a single-component object at radio frequencies, being detected by the Very Large Array (VLA), the Faint Images of the Radio Sky at Twenty--centimeters (FIRST, \citealt{Helfand2015}, the Very Large Array Sky Survey (VLASS, \citealt{Lacy2020}), LOFAR, and the Rapid ASKAP Continuum Survey (RACS, \citealt{Duchesne2025}).
The source spectrum is reasonably flat between 1 and 3 GHz, but with a peak emission in the LOFAR band. It is also fairly bright and extended in optical with a bright nucleus inside a dimmer envelope; its optical and infrared magnitudes are $\sim$11 and $\sim$10 in the $B$ and $K$ band, respectively. The source \emph{WISE} magnitudes are similarly fairly bright, but its far-infrared colours are atypical for highly excited AGN, being $W1-W2=0.14$, $W2-W3=1.85$, and $W3-W4=1.88$. 
Using the prescription of \citet{DAbrusco2019}, who used 3-dimensional \emph{WISE} colour space diagram to characterise AGNs/blazars 
(see Figure 6 in \citealt{DAbrusco2019}), one can conclude that these far-infrared colours are unusual for blazars and more typical of elliptical galaxies. 
Furthermore, these values suggest that the source might be either a low-excitation radio galaxy or a heavily absorbed AGN (\citealt{Dabhade2020}; \citealt{Gandhi2015}). However, the X-ray spectrum indicates that the intrinsic absorption is low, thus excluding this second scenario and pointing to a low excitation object.
Indeed, the DESI (Dark Energy Spectroscopic Instrument, \citealt{DESIColl2024}) spectrum shown in the Appendix, suggests an optical classification as a type 2 AGN, 
most likely a LINER, thus excluding this source as a candidate UEHBL system.\\

Finally, SWIFT J1949.7--3636 is easily associated with 2MASX J19495127--3635239, a local galaxy classified as a LINER-type AGN (\citealt{Mauch2007}); a more detailed analysis of the optical spectrum suggests it is indeed a type 2 AGN (see Appendix). Indeed, the X-ray spectrum, although of low statistical quality, suggests that some intrinsic absorption is present in the nucleus. The source is also radio detected by VLASS with a 3 GHz flux around 16 mJy and has \emph{WISE} colours ($W1-W2= 0.8$ and $W2-W3= 2.5$) typical of active galaxies. Overall, its properties confirm the active nature of its nucleus and, although its class remain somehow vague, also this object does not fit the requirements to be a UEHBL.

\subsection{More complex cases}

As anticipated, in 4 cases the association was not straightforward, and each of these sources is briefly discussed below to highlight their complexity.\\

\begin{figure}
	\centering 
\includegraphics[width=0.5\textwidth, angle=0]{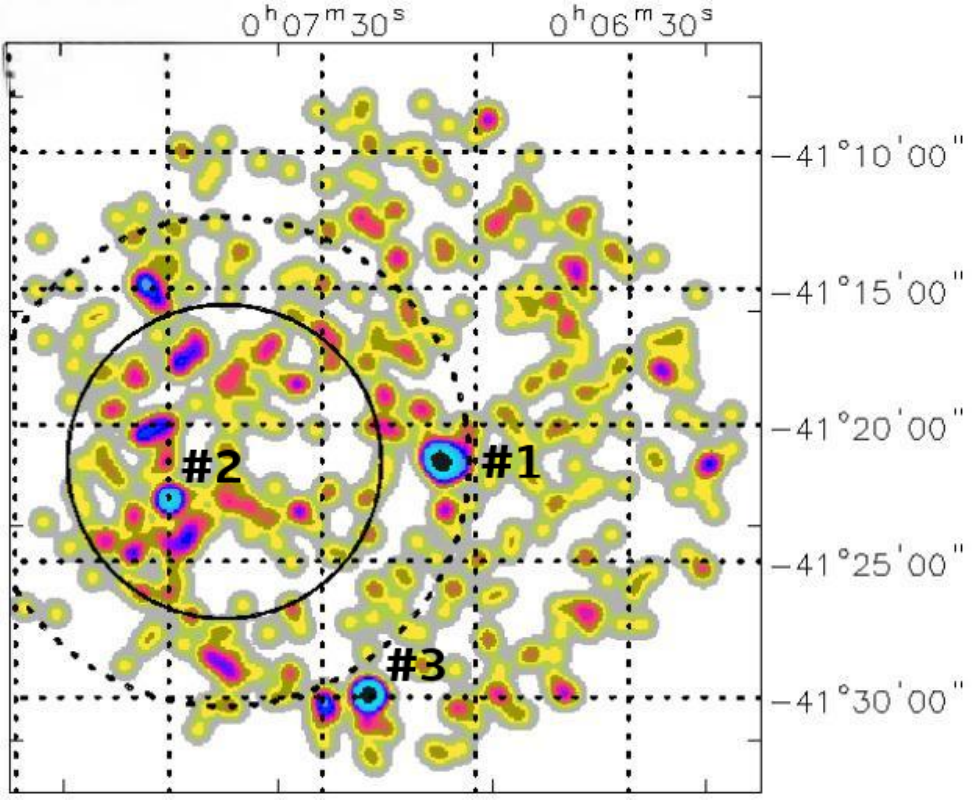}	
	\caption{XRT 0.3--10 keV image of the region surrounding 
    SWIFT J0007.8--4133.
    Source \#1 is the likely counterpart to the high-energy emitter (see text for details).}
	\label{fig1}
\end{figure}

In the case of SWIFT J0007.8--4133, the most plausible counterpart (source \#1) is outside the 90\% 157-month catalogue error circle but inside the 99\% one, as evident in Figure~\ref{fig1}; the only source (\#2) inside the BAT 157-month survey 90\% error circle is about 10 times weaker and also much softer, being not detected above 3 keV. Source \#3 is also undetected above few keV and in any case it is located outside the larger BAT positional uncertainty.
We therefore assume that source \#1 is the most likely counterpart. This X-ray source is associated with a relatively nearby galaxy named MCG--07--01--11/ESO 293--37, classified as a Seyfert 2 by \citet{Koss2022} based on data from the Southern Astrophysical Research telescope (SOAR) taken with the Goodman instrument; however, \citet{Zaw2019} suggest it maybe a type 1 AGN on the basis of its 6dF (The Six-degree Field Galaxy Survey, \citealt{Jones2009}) spectrum and the FWHM of the broad H$_\alpha$ line (1923 km s$^{-1}$). \citet{Chen2022} also classify the source as a Seyfert 1. 
Interestingly, we do not find evidence for intrinsic absorption in the X-ray spectrum as documented by the good fit to the data using a simple power law (see Table~\ref{tab2}), thus pointing again to a type 1 AGN. We thus adopt this last classification in Table~\ref{tab2}.
In any case, the presence of emission lines in the optical spectrum (whether broad or narrow) indicate that also this source cannot be considered as a valid UEHBL candidate anymore.\\

\begin{figure}
	\centering 
\includegraphics[width=0.5\textwidth, angle=0]{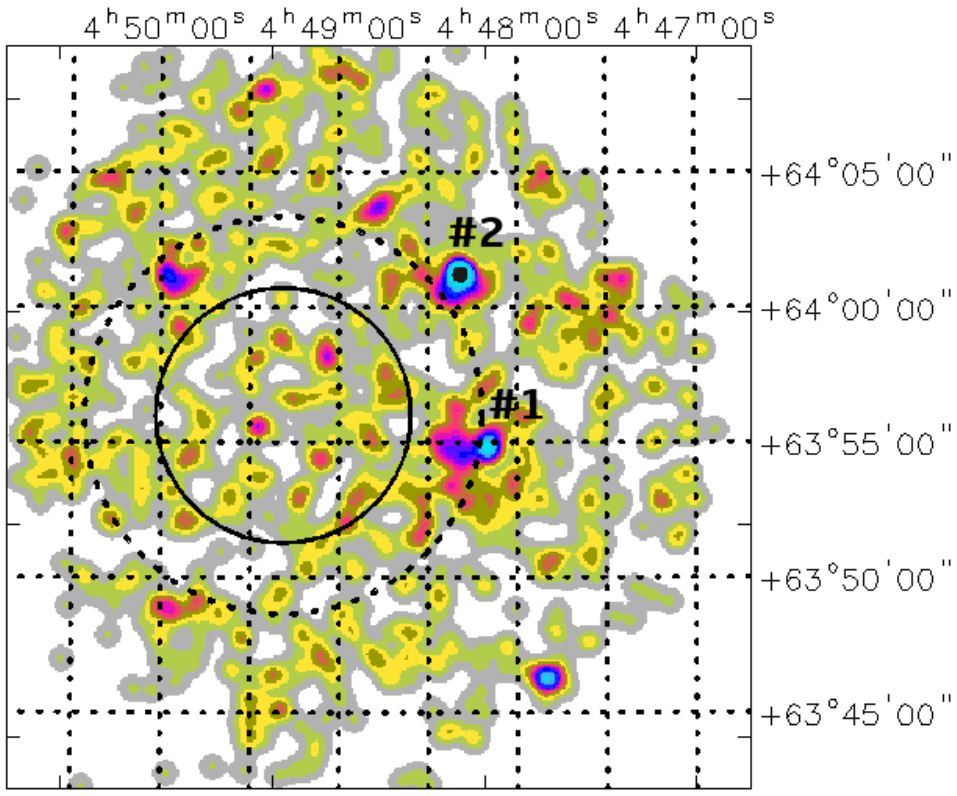}	
	\caption{XRT 0.3--10 keV image of the region surrounding SWIFT J0449.3$+$6356.
    Source \#1 and \#2 are the only X-ray detection found in this region (see text for details).}
	\label{fig2}
\end{figure}

SWIFT J0449.3$+$6356 is an interesting case as there is no evident X-ray detection within its 90\% positional uncertainty despite a more than 5-ks long exposure. The only two detections compatible with an enlarged error circle (99\% confidence level) are source \#1 at 7.7 arcmin distance from the BAT centroid and source \#2 at 8.4 arcmin distance (see Figure~\ref{fig2}). 
Their 2--10 keV fluxes are similar (see Table~\ref{tab2}), but both disappear above the few keV range, making both associations uncertain.

Source \#1 is probably associated with an optical/infrared source (WISE J044810.03$+$635447.2), with very little broad-band information. 

Source \#2 instead is most likely V* V374 Cam, an eclipsing binary candidate \citep{Kazarovets2011}, probably a detached Algol-type binary.
This Galactic source is nearby (266 pc), according to \citet{Bailer2021}, has an orbital period of 0.8 days, an  absolute optical magnitude $G$ = 4.6 and an X-ray to optical flux ratio around 10$^{-3}$; its \emph{Gaia} $BP-RP$ colour value is around 0.7. Following the prescription of \citet{Rodriguez2024}, it is possible to locate the source in the X-ray Main Sequence diagram using the above two parameters; this diagnostic diagram can be used to distinguish between accreting Galactic compact objects (containing either a white dwarf, a neutron star or a black hole) able to emit up to hard X-ray energies and active stars, which generally have only a soft high-energy spectral component. The source location is marginally compatible with the location of accreting compact objects, suggesting that it may be a possible counterpart to the BAT source despite its low X-ray luminosity 
($4\times10^{30}$ erg s$^{-1}$). Furthermore, the \emph{Gaia} magnitude versus colour diagram indicates that this optically variable source could also be a cataclysmic variable (CV) of the intermediate polar (IP) type \citep{Abril2020}, a class of Galactic objects often displaying 20--100 keV emission. However, this association remains tentative as the source is weak and outside the hard X-ray positional uncertainty; therefore, we still consider SWIFT J0449.3$+$6356 as a potential UEHBL candidate and discuss this case further in the next section.\\

\begin{figure}
	\centering 
\includegraphics[width=0.5\textwidth, angle=0]{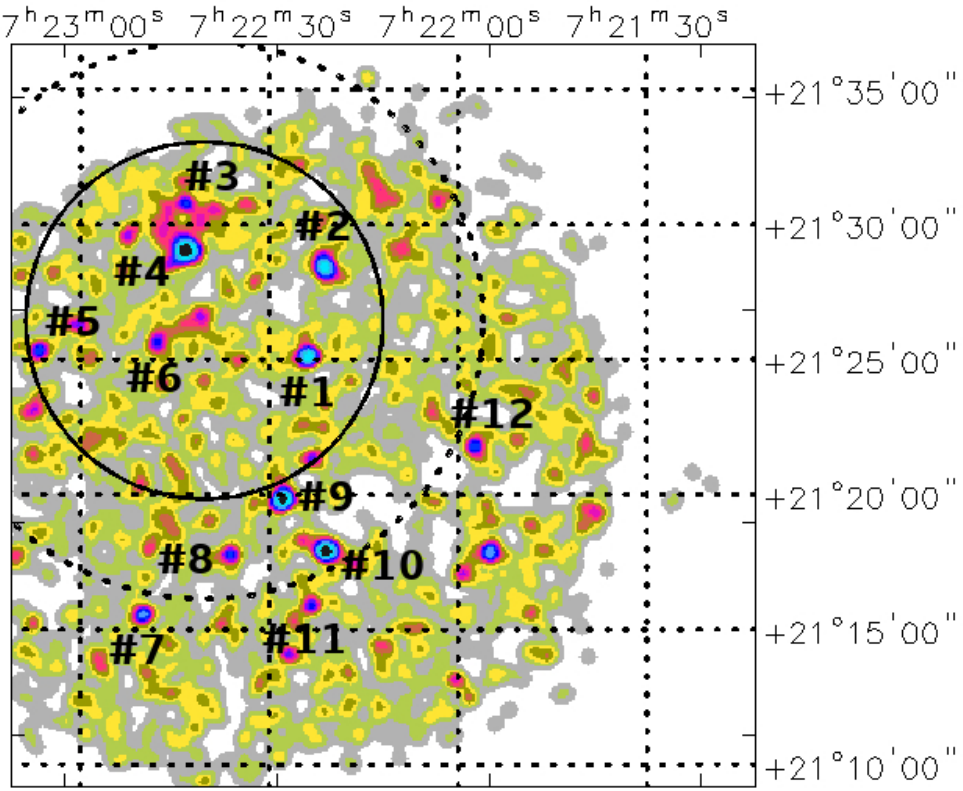}	
	\caption{XRT 0.3--2 keV image of the region surrounding 
    SWIFT J0722.5$+$2121.
    Source \#1 is the only X-ray detection found in this region above 2 keV (see text for details).}
	\label{fig3}
\end{figure}

On the opposite side is the case of SWIFT J0722.5$+$2121, where the long exposure (more than 12 ks) highlights a large number of X-ray detections both inside and outside its BAT positional uncertainty (see Figure~\ref{fig3}). However, these objects are all quite weak and no one is detected above 3 keV, although one source remains visible above 2 keV (source \#1 in Figure~\ref{fig3}); it is possibly associated with a weak
optical/infrared source (WISE J072223.57+212503.9), probably a background AGN at a photometric redshift $z=0.2$ \citep{Duncan2022}. Considering that it is the hardest source inside the BAT 90\% positional uncertainty, we assumed it to be the best possible association with SWIFT J0722.5$+$2121 and, as such, is reported in Table~\ref{tab1}. Given its relative weakness across most of the electromagnetic spectrum 
(characterised by a lack of radio detection, near-infrared brightness of $\sim$15 mag, optical magnitudes between 17 and 19, and the lowest X-ray flux in the sample, see Table~\ref{tab2}) it can still be considered as a viable extreme BL LAC candidate and therefore it is further discussed in the following section.\\

\begin{figure}
	\centering 
\includegraphics[width=0.5\textwidth, angle=0]{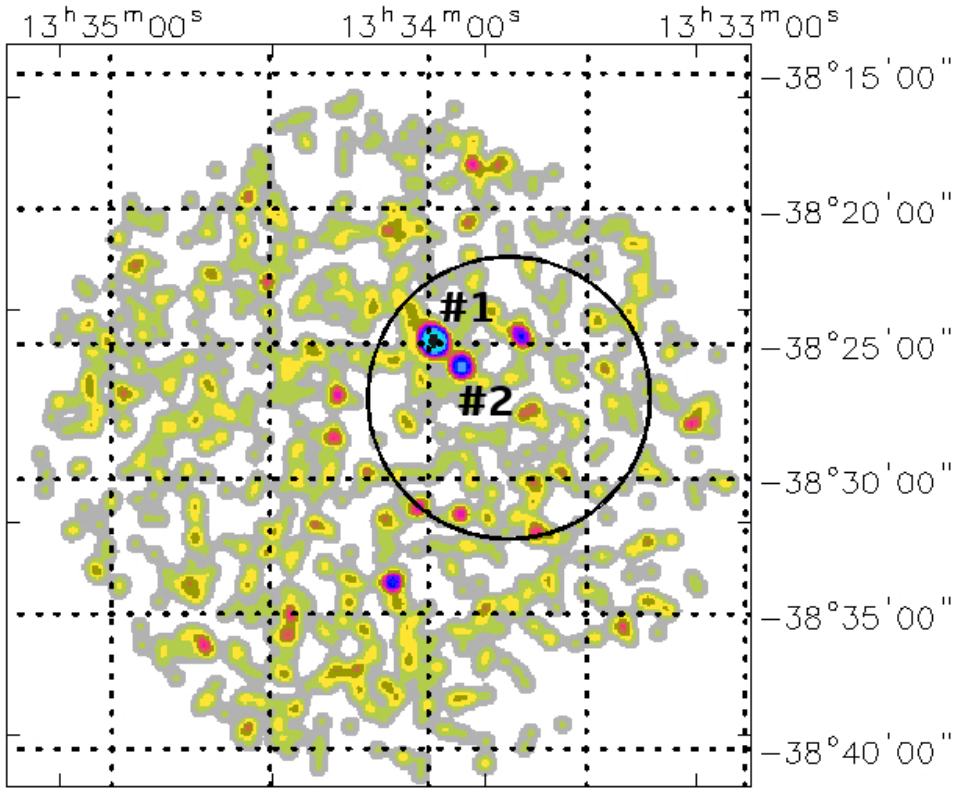}	
	\caption{XRT 0.3--10 keV image of the region surrounding 
    SWIFT J1334.1--3842.
    Source \#1 and \#2 are the only X-ray detection found in this region (see text for details).}
	\label{fig4}
\end{figure}

Lastly, in the case of SWIFT J1334.1--3842 the most likely counterpart is source \#1, which is associated with 2MASX J13335911--3824499, a Seyfert 1.9 galaxy of the nearby sky \citep{Koss2022}; the source maybe absorbed in X-rays, as expected for its optical spectral type, and has a hard X-ray spectrum with detection above 3 keV. 
However, we note another close by X-ray object (J133353.3--382548), indicated as source \#2 in Figure~\ref{fig4}, which is also quite hard being detected above a few keV. It is associated with LEDA 140189, a galaxy at a similar redshift of the Seyfert 1.9 galaxy above; its X-ray spectrum may also be absorbed, suggesting it might be a type 2 AGN. Indeed, its radio detection at a few mJy level by ASKAP and VLASS, as well as its infrared properties \citep{Edelson2012}, are suggestive of nuclear activity in the galaxy. This source is also interesting for being a luminous infrared galaxy detected by the InfraRed Astronomical Satellite (IRAS, see \citealt{Wang2009}) and 
AKARI \citep{Kilerci2018}; its infrared luminosity is $3.5\times10^{10}$ $L_{\odot}$.

As discussed in the Appendix, the source 6dF optical spectrum confirms the AGN nature of the source, but cannot discriminate between AGN types. 

Given its hardness at X-ray energies, we cannot exclude this source as a potential contributor to the hard X-ray emitter. In any case, neither source fits with the requirements needed to fulfill the definition of extreme BL Lac and therefore neither will be considered in the following discussion.

\begin{table*}
\begin{center}
\footnotesize
\caption{Counterparts of the 10 high-energy sources discussed in this work. For each of these candidate counterparts we report the optical name, classification and redshift (with corresponding reference), the Galactic/intrinsic X-ray column densities, the X-ray photon index, the $(\chi^{2}/C-stat)/d.o.f.$, and the 2--10 keV flux. Frozen parameters are written in square brackets; errors are given at the 90\% confidence level.}
\label{tab2}
\scriptsize
\begin{tabular}{lcccccccc}
\hline
\hline
  & &     &   &   &  & &  & \\
XRT source & Optical ID  &  $z$ & Ref. & $N_{H(Gal)}$ & $N_{H(int)}$ &
$\Gamma$ & $(\chi^{2}/C-stat)^{a,b}/d.o.f.$ & $Flux_{(2-10~keV)}$ \\
 & (class) &  &   &   & &   & \\
   
  &    & & &  (10$^{22}$ cm$^{-2}$) & (10$^{22}$ cm$^{-2}$) &  &  
&  (10$^{-12}$ erg cm$^{-2}$ s$^{-1}$)\\ 
 &          &   & &    & &   &   &    \\
\hline
\hline
\multicolumn{9}{c}{\textbf{SWIFT J0007.8--4133}}  \\
\#1$^{c}$ & MCG--07--01--011  &  0.047596 & 1 & 0.0131 & -- & $1.50\pm0.11$ & 31.3/29$^{a}$  & $1.49\pm0.06$  \\
   &  (Seyfert 1) &    &    &   &  &   &    \\
   \hline
      \multicolumn{9}{c}{\textbf{SWIFT J0045.9$+$3931}}  \\
\#1  & LEDA 2151989 & 0.077427 & 2/3 & 0.0463 &  $2.00^{+2.53}_{-0.88}$  & [1.8] &  9.6/7$^{a}$  &   $0.78\pm0.13$  \\
&   (Seyfert 2) &   &   &   &   &   &   &   \\
   \hline
      \multicolumn{9}{c}{\textbf{SWIFT J0106.1$+$4818}} \\
\#1$^{d}$  &  ICRF J010549.9$+$481903  & -- & 4 & 0.125 & $2.70^{+1.27}_{-0.86}$  & [1.8] &  7.9/8$^{a}$ &   $1.43\pm0.14$ \\
&   (Blazar candidate) &   &   &   &   &   &   &   \\
   \hline
      \multicolumn{9}{c}{\textbf{SWIFT J0243.2--0553}}  \\
 \#1 & QSO B0240--060  & 1.805 & 5 & 0.0264 & -- & $1.67\pm0.51$ & 20.1/22$^{b}$   &  $0.95\pm0.20 $ \\
 &   (FSRQ) &   &   &   &   &   &   &   \\
   \hline
      \multicolumn{9}{c}{\textbf{SWIFT J0449.3$+$6356 }}\\
\#1  & WISE J044810.03+635447.1  & -- & 2 & 0.193 &--  & $1.31^{+0.66}_{-0.67}$  & 19.7/18$^{b}$ &  $0.36\pm0.10$  \\
&   (AGN?) &   &   &   &   &   &   &   \\
\#2  & V* V374 Cam  &--  &  4 & 0.194$^{e}$ & -- & $2.07\pm0.65$  &  22.3/20$^{b}$ & $0.22\pm0.04$   \\
&   (Eclipsing Binary) &   &   &   &   &   &   &   \\
   \hline
      \multicolumn{9}{c}{\textbf{SWIFT J0656.0--6560}}\\
  &  &     &  &   & &  &  &  \\
  \#1$^{f}$ & FRL 265  & 0.030451 & 4 & 0.058 & -- & $1.97\pm0.08$ & 76.3/59$^{a}$  & $1.63\pm0.04$  \\
&   (Seyfert 1) &   &   &   &   &   &   &   \\
   \hline
      \multicolumn{9}{c}{\textbf{SWIFT J0722.5$+$2121}}\\
  \#1 & WISE J072223.57+212503.9 & 0.194 &  2    & 0.0747 & -- & $0.90^{+0.71}_{-0.77}$ &  21.5/24$^{b}$ & $0.13\pm0.04$  \\
&   (AGN?) &   &   &   &   &   &   &   \\
   \hline
      \multicolumn{9}{c}{\textbf{SWIFT J1026.3$+$4536}}\\
\#1   & Z 240--42  & 0.02689 & 4/2 &  0.0152  &   $1.40^{+1.00}_{-0.79}$  & $2.06^{+0.87}_{-0.79}$ & 8.9/8$^{a}$ &  $0.43\pm0.04$\\
&   (Type 2 AGN) &   &   &   &   &   &   &   \\
   \hline
      \multicolumn{9}{c}{\textbf{SWIFT J1334.1--3842}}  \\
 \#1$^{g}$ & 2MASX J13335911--3824499 &  0.05333 & 6 &  0.0445 &  $2.94^{+2.03}_{-1.32}$  & [1.8] & 5.0/8$^{a}$ & $1.80\pm0.27$ \\
&   (Seyfert 1.9) &   &   &   &   &   &   &   \\
 \#2  & LEDA 140189   & 0.05072 & 2 & 0.0447 & $9.81^{+19.70}_{-5.72}$ & [1.8] & 15.8/16$^{b}$ & $0.76\pm0.20$  \\
&   (Type 2 AGN) &   &   &   &   &   &   &   \\
   \hline
      \multicolumn{9}{c}{\textbf{SWIFT J1949.7--3636}}  \\
\#1 &  2MASX J19495127--3635239  & 0.046746 & 4/2 & 0.0774  & $6.12^{+1.39}_{-1.04}$ & [1.8]  & 13.8/15$^{a}$ & $3.41\pm0.17$\\
&   (Type 2 AGN) &   &   &   &   &   &   &   \\
\hline
\hline
\end{tabular}
\begin{list}{}{}
\item $^{a}$: For this source we applied the $\chi^2$ statistics;
\item $^{b}$: For this source we applied the Cash statistic;
\item $^{c}$: The source is also reported in the \emph{XMM-Newton} Slew catalogue as 
XMMSL3 J000706.7--412118 with a 0.2--12 keV of $3.0\times10^{-12}$ erg cm$^{-2}$ s$^{-1}$; 
\item $^{d}$: The source is also reported in the \emph{XMM-Newton} Slew catalogue as XMMSL3 J010549.1$+$481908 with a 0.2--12 keV flux in the range $(1.4-3.2)\times10^{-12}$ erg cm$^{-2}$ s$^{-1}$;
\item $^{e}$: For this source the Galactic column density is not required by the data, probably due to its proximity;
\item $^{f}$: The source is also reported in the  
\emph{XMM-Newton} Slew catalogue as XMMSL3 J065630.5--653336 with a 0.2--12 keV flux in the range $(0.2-1.2)\times10^{-12}$ erg cm$^{-2}$ s$^{-1}$;  
\item $^{g}$: The source is also reported in the eROSITA/eRASS1 hard catalogue
as eRASS J133359.2--382449  with a 2.3--5.0 keV flux of $7.4\times10^{-13}$ erg cm$^{-2}$ s$^{-1}$;
\item References to optical data: (1) \citet{Zaw2019}; (2) this work (see Appendix); 
(3) \citet{Wang2018}; (4) SIMBAD database; (5) \citet{Hook2003}; (6) \citet{Koss2022}.
\end{list}
\end{center}
\end{table*}

\section{Surviving UEHBLs candidates}

As evident from Table~\ref{tab2}, a number of associations are not BL Lac objects but rather AGNs showing emission and/or absorption line spectra; as their optical characteristics do not fit with a classical BL Lac classification, they can be easily taken out from the list of UEHBL candidates as described in detail in the previous section. 

The only surviving candidates in our sample are 3 objects: SWIFT J0106.1$+$4818, associated with the AGN/blazar ICRF J010549.9$+$481903 and the 2 hard X-ray emitters SWIFT J0449.3$+$6356 and SWIFT J0722.5$+$2121, for which we were not able to find a convincing X-ray/optical counterpart, but for which some X-ray detections could prove to be interesting sources to explore further.

In the following, we describe each of these individual cases separately. 

\subsection{SWIFT J0106.1$+$4818/ICRF J010549.9$+$481903}

The counterpart of this hard X-ray source, ICRF J010549.9$+$481903, is classified as a compact flat spectrum radio source well studied over the radio wavebands: the core is 0.2 mas in size \citep{Dodson2008} and two more components can be traced on opposite sides of the core \citep{Britzen2007}. Although the source spectral index is flat over the 1--5 GHz range, \citet{Marecki1999} noted that this is a Gigahertz-Peaked-Spectrum object, as the overall spectral shape up to 43 GHz shows a peak at a few GHz.

\begin{figure}
	\centering 
\includegraphics[width=0.5\textwidth, height=0.3\textwidth, angle=0]{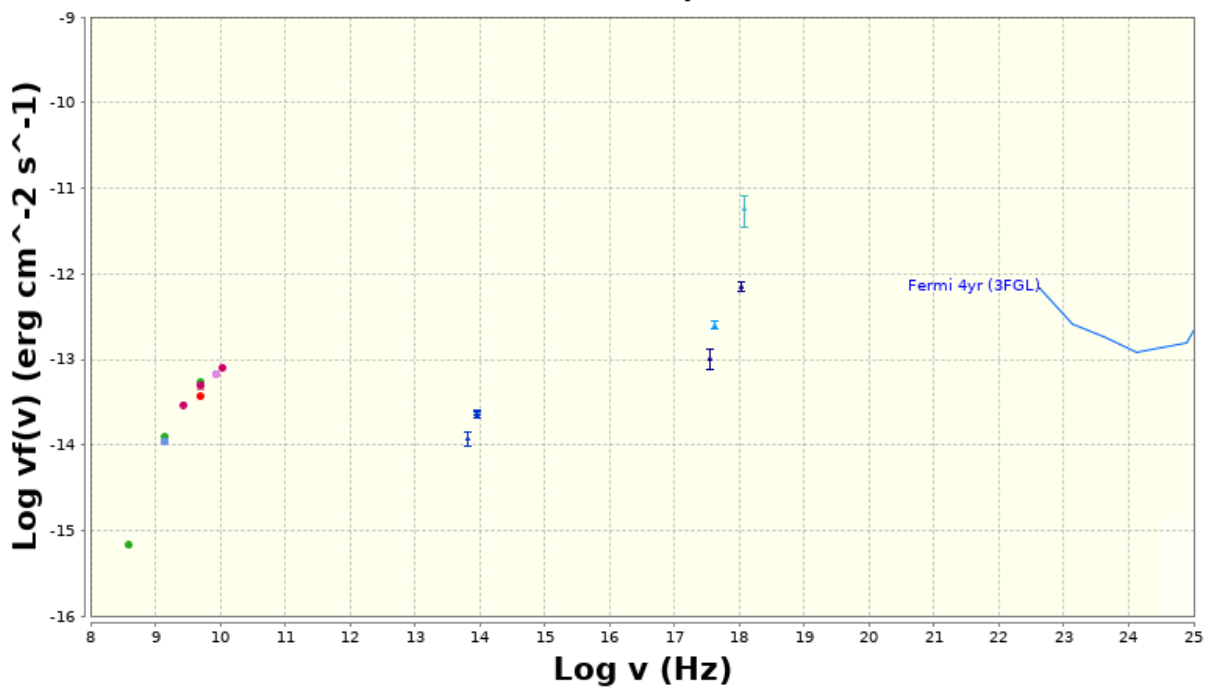}	
	\caption{SED of ICRF J010549.9$+$48190. SED measurements are from archival data from the SSDC database and current XRT analysis. See text for details.}
	\label{fig5}
\end{figure}

The source has been monitored over the years, as it shows strong variability at various frequencies (see for example \citealt{Richards2014} and \citealt{Liodakis2017}), and its emission is also polarised \citep{Jackson2007}. All these properties point to a blazar nature for the source, but no optical spectrum is available to confirm this suggestion. The source is in fact quite weak in the infrared domain (magnitudes around 17--18 in the \emph{WISE} band) and even weaker, around 24 mag, in the $R$ band \citep{Healey2008}, hence the difficulty to obtain an optical classification for the source.

So far, the \emph{Fermi}/LAT instrument has not detected the source as a gamma-ray emitter, allowing for the possibility that this could be a MeV-peaked blazar. 
Assuming instead that this source is an ultra extreme high-energy-peaked BL Lac, we note that the broadband X-ray spectrum is compatible with case B/C of the SED modeling discussed by \citet{Sciaccaluga2025} in their paper
(see source SED in Figure~\ref{fig5} as obtained from the ASI ASDC SED Builder page at https://tools.ssdc.asi.it/SED/); broadly, also the infrared and optical data are in agreement with models B/C, assuming fluxes around $2\times10^{-14}$ and $4\times10^{-15}$ erg cm$^{-2}$ s$^{-1}$ at 10$^{14}$ and $5\times10^{14}$ Hz, respectively. It remains to understand why the radio emission is so bright with respect to other wavebands, unless variability or an extra component explain this discrepancy in the source SED. Ideally, only an optical spectrum could clarify the nature of this object and thus confirm its extreme BL Lac nature, but given the source weakness we find such measurement challenging.

\subsection{SWIFT J0449.3$+$6356/WISE J044810.03+635447.1}

Keeping in mind that an alternative Galactic association is possible, we explore a scenario where either source \#1 or some other still undetected object inside the BAT positional uncertainty is the possible extragalactic counterpart to SWIFT J0449.3$+$6356; in either case, the properties of object \#1 can be taken as indicative of the spectral behavior or SED of this high-energy emitter.

WISE J044810.03$+$635447.1 is a poorly studied source: no radio detection is so far available, nor a photometric redshift has been estimated. The only available data are \emph{WISE} magnitudes and colours, which suggest this may be a weak extragalactic AGN: fluxes are around $(1-2)\times10^{-13}$ erg cm$^{-2}$ s$^{-1}$, while $W1-W2$ and $W2-W3$ are 0.5 and 2.9 mag, respectively. The only other available information came from Panoramic Survey Telescope and Rapid Response System (Pan--STARRS) data (see \citealt{Magnier2020} and subsequent updates), reporting optical fluxes around $(0.5-1.6)\times10^{-13}$ erg cm$^{-2}$ s$^{-1}$. These values are broadly compatible with case A/B of the SED modeling discussed by \citet{Sciaccaluga2025} in their paper, although the X-ray data fit case C better. Considering source variability and poor data coverage, this discrepancy is acceptable and the \emph{WISE} source can still be considered a valid UEHBL candidate. In any case, further monitoring by \emph{Swift}/XRT of this sky region can provide useful information regarding source variability and further evidence for its association with the BAT emitter.

\subsection{SWIFT J0722.5$+$2121/WISE J072223.57$+$212503.9} 

As anticipated, this source is very weak in X-rays, as well as over most of the electromagnetic spectrum.
Its optical magnitude ranges from 18--19 in the $B$ band to 17 in the $R$ one, reaching 21 magnitude in the UVW1 filter as seen with \emph{Swift}/UVOT; similarly, in the near-infrared, the magnitudes in the $J$ to $H$ wavebands are in the range 15--16. In the far-infrared, the \emph{WISE} magnitudes are also quite dim (around 14--15 in the two shortest \emph{WISE} wavebands), while the relative colours are just compatible with an AGN nature ($W1-W2=0.4$ and $W2-W3=3.3$). The source has recently been detected at radio frequencies since it is reported in the LOFAR DR3 catalogue with a flux of 0.2 mJy at 144 MHz \citep{Shimwell2026}, while at few GHz its flux is below the mJy level according to the VLASS data survey \citep{Gordon2021}.

\begin{figure}
	\centering 
\includegraphics[width=0.5\textwidth, height=0.3\textwidth, angle=0]{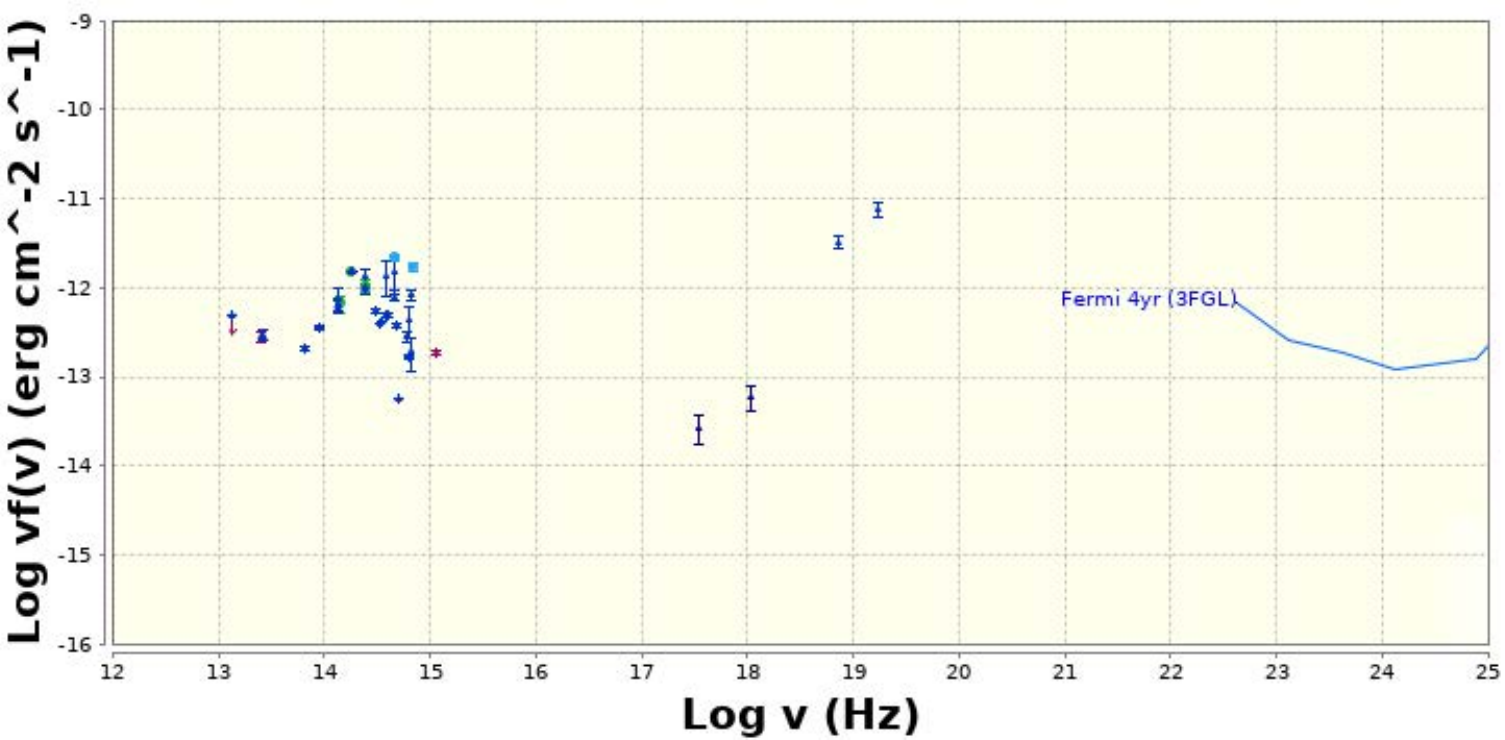}	
	\caption{SED of WISE J072223.57$+$212503.9. 
    SED measurements are from archival data from the SSDC database and current XRT analysis.
    See text for details.}
	\label{fig6}
\end{figure}

Although these numbers convert to weak fluxes over the radio to X-ray range, as expected in the case of extreme high-energy-peaked BL Lacs, the SED shape (as obtained from the ASI ASDC SED Builder page at https://tools.ssdc.asi.it/SED/) is reminiscent of a more classical blazar type AGN 
(see Figure~\ref{fig6}), with the synchrotron peak in the optical/infrared waveband and, consequently, the Compton peak at MeV frequencies. If WISE J072223.57$+$212503.9 is the real counterpart to this BAT source, it is most likely a classical quasi-stellar object (QSO) at a relatively nearby distance (photometric redshift around 0.2 from \citealt{Beck2022}). Again, only an optical spectrum of the source will define its final nature, but from the available data it can already be excluded from the list of UEHBL candidates.


\section{Summary and Conclusions}

We explored the \emph{Swift}/XRT data of a sample of 10 still unidentified hard X-ray sources extracted from the \emph{Swift}/BAT 157-month survey catalogue, which have been proposed as the first examples of a population of ultra extreme high-energy-peaked BL Lacs, i.e. objects with a synchrotron peak in the MeV waveband.
We have been able to firmly associate 7 of these sources to line-emitting AGN, thus excluding a BL Lac nature and then an extreme blazar nature. As a byproduct of this analysis, we have been able to classify 3 of these objects 
(SWIFT J0007.8--4133, SWIFT J0045.9$+$3931, and SWIFT J1949.7--3636) as type 2 AGN and 2 more (SWIFT J0243.2--0553 and SWIFT J0656.0--6560) as broad line AGN. In one case, that of SWIFT J1334.1--3842, we find that the hard X-ray emission is probably the contribution of two absorbed AGNs, i.e. a Seyfert of type 1.9 and another of type 2. 
Lastly, SWIFT J1026.3$+$4536 is associated with a low excitation radio galaxy, likely of the LINER subclass.

Of the remaining 3 UEHBL candidates, only 2 objects are found to have a SED compatible with those expected by MeV-peaked BL Lacs: SWIFT J0106.1$+$4818 and SWIFT J0449.3$+$6356. However, in this last case, an alternative association with a Galactic source is also possible. 
SWIFT J0722.5$+$2121 is instead most likely associated with a classical QSO.

Overall, we find that if ultra extreme high-energy-peaked BL Lacs do really exist, they are extremely rare to find and furthermore difficult to characterise and properly classify. In any case, they represent a valid association to explore for those hard X-ray emitters which have no obvious X-ray counterpart, either because weak or non-detected.

\section*{Acknowledgements} 
This research has made use of data obtained from the VizieR catalog access tool and the SIMBAD database, which are both operated at CDS, Strasbourg, France; the High Energy Astrophysics Science Archive Research Center (HEASARC), a service of the Astrophysics Science Division at NASA/GSFC.
We also acknowledge the use of public data from the \emph{Swift} data archive and the Space Science data Center SED Builder tool at https://tools.ssdc.asi.it/SED/.

\section{Appendix}

In the following, we briefly give more details on the classification of the 4 sources illustrated in the main text on the basis of available archival optical spectra
using the BPT diagram prescription \citep{Baldwin1981} as reference.

\begin{figure}
	\centering 
\includegraphics[width=0.5\textwidth, angle=0]{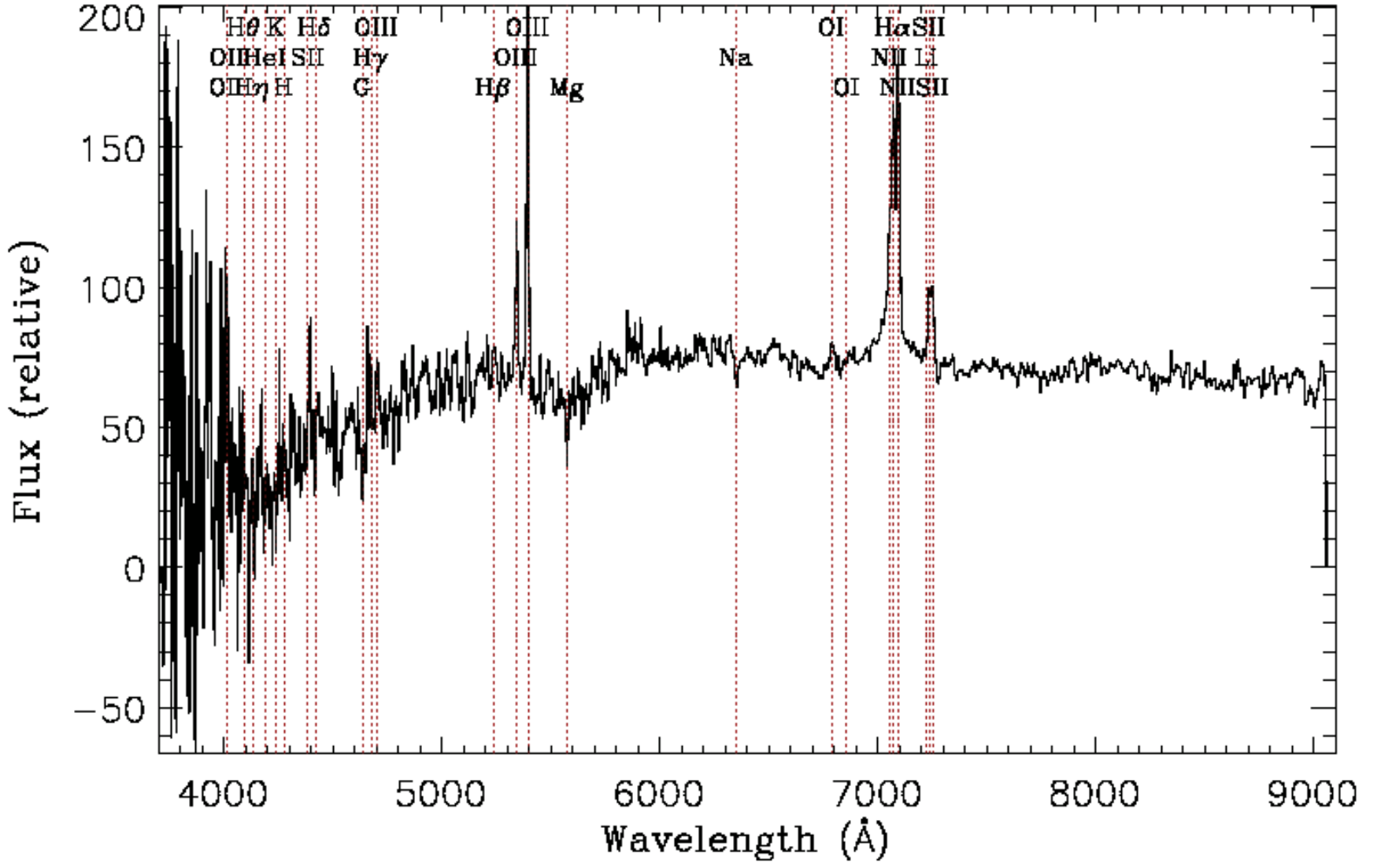}	
	\caption{LAMOST spectrum of J004558.1$+$393352.5/LEDA 2151989 \citep{Cui2012}.}
	\label{A1}
\end{figure}

\begin{figure}
	\centering 
\includegraphics[height=0.3\textwidth, width=0.5\textwidth, angle=0]{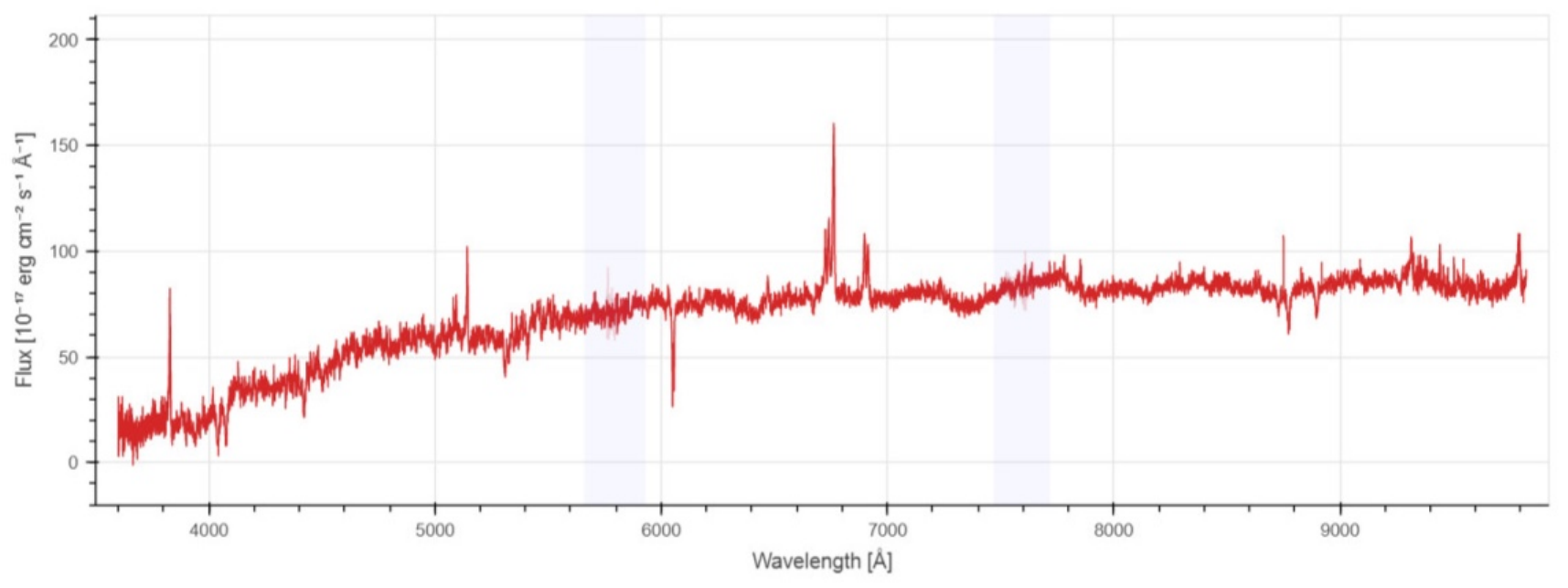}	
	\caption{DESI spectrum of J102618.9$+$453445/Z 240--42 \citep{DESIColl2024}.}
	\label{A2}
\end{figure}

\begin{figure}
	\centering 
\includegraphics[width=0.49\textwidth, angle=0]{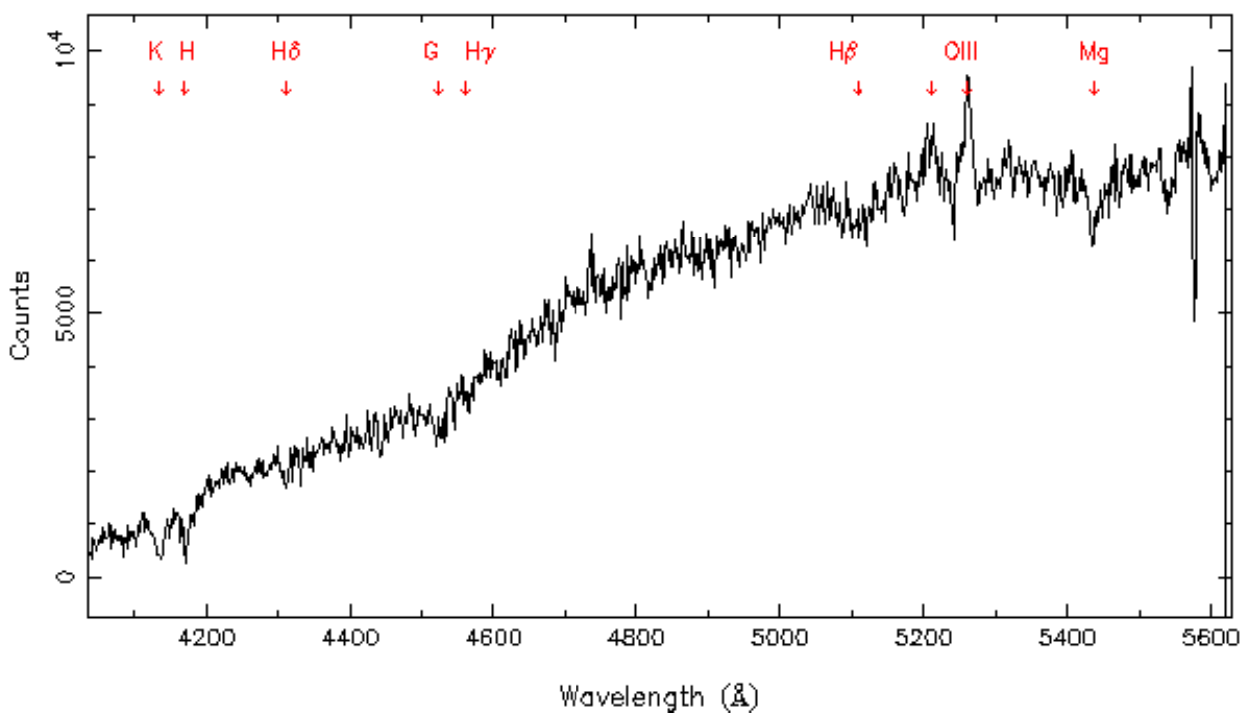}	
\includegraphics[width=0.49\textwidth, angle=0]{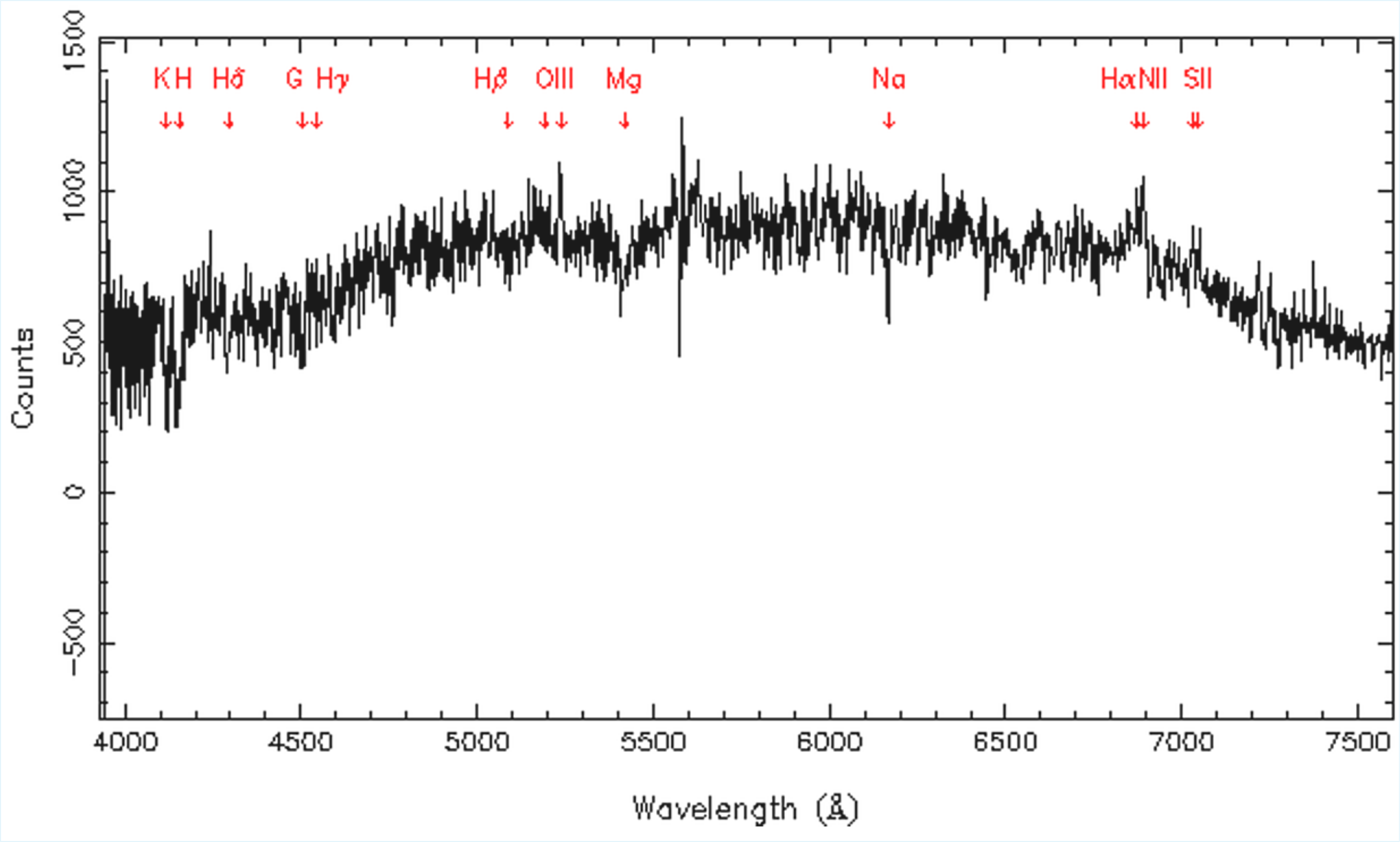}	
	\caption{6df spectrum \citep{Jones2009} of J133353.3--382548/LEDA 140189 (upper panel) and J194951.1--363522/2MASX J19455127--3635239 (lower panel).}
	\label{A3}
\end{figure}

\subsection{LEDA 2151989}

J004558.1$+$393352.5, also named LEDA 2151989, is classified as a Seyfert galaxy at redshift $z = 0.077427$ by \citet{Wang2018} based on the spectrum obtained by the Large Sky Area Multi-Object Fiber Spectroscopic Telescope (LAMOST, \citealt{Cui2012}) since its line flux ratios in logarithmic scale are [O{\sc iii}]/H$_\beta$ = 0.96 and [N{\sc ii}]/H$_\alpha$ = 0.13. A quick look at the LAMOST spectrum further allows a more precise classification as a Seyfert of type 2 given the narrowness of all its emission lines, both permitted and forbidden (see Figure~\ref{A1}).

\subsection{J102618.9$+$453445}

This source, also named Z 240--42, is classified in SIMBAD as a relatively closeby ($z = 0.0267$) radio galaxy. Actually, its DESI spectrum (see Figure~\ref{A2}) is characterised by the presence of several narrow permitted and forbidden emission lines at the above redshift, with [N{\sc ii}] stronger than H$_\alpha$ by a factor of $\sim$2, a barely detectable H$_\beta$, and [O{\sc ii}] and [O{\sc iii}] emission of comparable strength.
This information allows us to rule out a BL Lac nature for this source; rather, it can be classified as a type 2 AGN, most likely of the LINER subclass.

\subsection{J133353.3$-$382548}    
The available 6dF spectrum of this source, also labelled as LEDA 140189, has a narrow coverage (4000--5600 \AA) and provides a redshift of $z = (0.05072\pm0.00018)$ (\citealt{Jones2009}); the  continuum is typical of a red galaxy with  no apparent emission lines other than the [O {\sc iii}] doublet (see Figure~\ref{A3}, upper panel). From this spectrum one can thus say that this may be an active galaxy, but more overall information and more spectral coverage are needed. We note, however, that the source redshift is similar to that of 2MASX J13335911--3824499, suggesting that both may belong to the same group of galaxies.

\subsection{J194951.1$-$363522}

According to \cite{Mauch2007}, this galaxy (also named 2MASX J19495127--3635239, at redshift $z = (0.04676\pm0.00015$) shows narrow LINER-like emission lines. However, its 6dF spectrum (see Figure~\ref{A3}, lower panel), albeit noisy, does not allow us to confirm this classification due to the fact that the [O {\sc ii}] falls bluewards of the covered spectral range (4000--7500 \AA); see \citet{Heckman1980} for the definition of LINER. To be safe, one can state that the object resembles a type 2 AGN due to the presence of H$_\alpha$, [N {\sc ii}], [O {\sc iii}], and [S {\sc ii}] narrow emission features and the absence of H$_\beta$ emission.

\bibliographystyle{elsarticle-harv} 
\bibliography{reference}






\end{document}